\documentclass[12pt]{article}

\usepackage{scicite}
\usepackage{graphicx}
\usepackage{xr}
\usepackage{times}
\usepackage{array}
\usepackage{url}
\usepackage{color}
\newcolumntype{P}[1]{>{\centering\arraybackslash}p{#1}}
\newcolumntype{M}[1]{>{\centering\arraybackslash}m{#1}}
\DeclareMathAlphabet{\mathpzc}{OT1}{pzc}{m}{it}

\newenvironment{sciabstract}{%
\begin{quote} \bf}
{\end{quote}}

\title{Urbanization affects peak timing, prevalence, and bimodality of influenza pandemics in Australia: results of a census-calibrated model}

\author
{Cameron Zachreson,$^{1\ast}$ Kristopher M. Fair,$^{1}$ Oliver M. Cliff,$^{1}$\\ Nathan Harding,$^{1}$ Mahendra Piraveenan,$^{1}$ Mikhail Prokopenko$^{1,2}$\\
\\
\normalsize{$^{1}$Complex Systems Research Group, School of Civil Engineering,}\\
\normalsize{Faculty of Engineering and IT, The University of Sydney,}\\
\normalsize{Sydney, New South Wales, Australia, 2006.}\\
\normalsize{$^{2}$Marie Bashir Institute for Infectious Diseases and Biosecurity,}\\
\normalsize{University of Sydney, Westmead, NSW 2145, Australia.}\\
\\
\normalsize{$^\ast$To whom correspondence should be addressed; e-mail:  cameron.zachreson@sydney.edu.au}
}

\date{}

\begin{document} 


\baselineskip24pt


\maketitle

\section{One Sentence Summary:}
Simulations associate urbanization with earlier peaks, higher peak prevalence, and shifting bimodality of pandemics in Australia.

\pagebreak
\section{Abstract:}
\begin{sciabstract}
We examine salient trends of influenza pandemics in Australia, a rapidly urbanizing nation. To do so, we implement state-of-the-art influenza transmission and progression models within a large-scale stochastic computer simulation, generated using comprehensive Australian census datasets from 2006, 2011, and 2016. Our results offer a simulation-based investigation of a population's sensitivity to pandemics across multiple historical time points, and highlight three significant trends in pandemic patterns over the years: increased peak prevalence, faster spreading rates, and decreasing spatiotemporal bimodality. We attribute these pandemic trends to increases in two key quantities indicative of urbanization: population fraction residing in major cities, and international air traffic. In addition, we identify features of the pandemic's geographic spread that we attribute to changes in the commuter mobility network. The generic nature of our model and the ubiquity of urbanization trends around the world make it likely for our results to be applicable in other rapidly urbanizing nations.
\end{sciabstract}
\pagebreak

\section*{Main Text:}
\subsection{introduction}
The global population is both highly urbanized, and rapidly urbanizing, with people around the world flocking to cities more quickly every year since 1950 \cite{nations2014world}. This critical trend has inspired a significant research effort towards understanding the economic, environmental, and social implications of global urbanization \cite{Wigginton904,Seto943,Bloom772}. To further this effort, we present findings on the relationship between urbanization trends and pandemic influenza sensitivity. 

In general, urbanization is known to worsen epidemics through a variety of mechanisms. In developing nations this is well-documented and is associated with high-density living conditions combined with concentrated poverty and poor sanitation \cite{world2008our}. In the developed world, where urban living has been the norm for the last century, the direct effects of further population concentration into cities on potential pandemic dynamics are less obvious. The concentration of healthcare facilities, combined with increased economic opportunities tends to increase the general quality of public health \cite{dye2008health}. However, with respect to communicable disease, the concentration of the workforce in central business districts, combined with suburban sprawl, can produce large hubs in the commuter interaction network that could potentially lead to faster proliferation of infectious disease between work and home \cite{eubank2004modelling,yashima2014epidemic}. Additionally, city growth is coupled to increased air traffic and connectivity, which could be expected to increase both the probability of inter-city disease spread, as well as the potential for the disease to arrive from overseas due to international traffic \cite{brueckner2003airline}.

Our study focuses on Australia, a highly urbanized nation for which comprehensive census data spanning the last decade has been made publicly available. We used this data to calibrate a nation-level model of pandemic influenza spread, and investigate the surrogate population's vulnerability to the contagion over a period of rapid urbanization. 

Australia has a total urban population fraction of  about $90 \%$, with over half of the country's population located within only a few urban centers. Over time, this trend has led to significant  infrastructural stress, including increased healthcare economic burden. In particular, the cost of influenza in Australia is estimated to be between \$828 and \$884 million annually \cite{newall2007economic}. Rapidly escalating epidemics directly impact other aspects of life. Individual behavior changes, and the resilience of critical infrastructure is called into question \cite{o2007critical,boin2007preparing}. Impacts are quickly seen also in labor supply and productivity, in the movement of goods and services across regions, and in the sustainability of consumer and investor confidence. Arguably the most acute aspect of an epidemic or pandemic crisis is the strain on medical infrastructure. In the last decade, Australia's hospital beds have been maintained at approximately 2.5 per 1000 individuals, leading major hospitals to regularly operate from 90$\%$ to 100+$\%$ capacity. The Australian Medical Association has assessed the condition as unacceptable, but hospital capacity continues to lag behind demand \cite{australian2010public}. The constant (low) per capita bed numbers and the country's aging population make any relative increases in disease prevalence significant from the perspective of hospital over-crowding and the related adverse health outcomes for patients and staff, including increased patient mortality \cite{sprivulis2006association}.

Despite its relative isolation, Australia was not spared from the H1N1 influenza (``swine flu'') pandemic in 2009 \cite{bishop2009australia}. Since then, the prevalence of seasonal influenza in Australia has been increasing on average. Indeed, the country experienced a particularly severe season in 2017, with levels approaching those of 2009 \cite{flureport}. To make matters worse, a recent analysis has assessed Australia's pandemic preparedness as sorely lacking \cite{itzwerth2018australia}. Based on these general trends we can expect future pandemics to be more damaging.

Our motivation for the present work is the prospect of linking long-term structural and social trends to Australia's apparently increasing susceptibility to contagion (in this case, the influenza virus). 

%There are many existing methods and tools for modeling epidemics, as well as assisting in crisis response and preparedness planning \cite{nsoesie2012sensitivity,nsoesie2014systematic,eubank2004modelling,barrett2008episimdemics}. 
There are many existing methods and tools for modeling epidemics, as well as assisting in crisis response and preparedness planning \cite{nsoesie2014systematic,eubank2004modelling}. However, none have been used to provide a high-resolution comparative analysis of pandemic trends across multiple historical timepoints. Here, we accomplish this by simulating the spread of influenza through a stochastically-generated population mimicking the Australian Bureau of Statistics (ABS) censuses of 2006, 2011, and 2016.  We conduct our simulations at the community- and national-level in a way that rigorously accounts for salient features of the contagion, such as prevalence dynamics and spatiotemporal structure. Importantly, this approach allows us to investigate the role of the slowly evolving population distribution and social interaction network over which the virus spreads, independently of the viral characteristics. Such a separation of contagion and population properties is impossible in field studies, because both are intrinsically dynamic \cite{petrova2018evolution}. 

Our results highlight three significant trends in epidemic patterns over the years, which are independent of the specific simulation parameters defining the influenza virus:

\begin{itemize}
\item{Increased peak prevalence}
\item{Faster spreading rates (earlier epidemic peak)}
\item{Decreasing spatiotemporal bimodality}
\end{itemize}

These results have important implications: the first point predicts increased intensity of health crises in infrastructural terms, with more people simultaneously expressing flu symptoms during the peak of the epidemic; the second point indicates shorter periods during which detection and response strategies would have to be implemented, for effective mitigation; the third point pertains to the bimodal character of the epidemic which occurs in two waves, the first in cities near international airports (where the pandemic is introduced), and the second in rural areas and cities without international airports. In this case, decreasing bimodality occurs as the first mode subsumes the second, contributing to the severity of the epidemic at its peak. 

Bimodal H1N1 outbreaks were observed around the world during the 2009 global H1N1 pandemic \cite{chowell2011spatial,choudhry2012emergence,echevarria2009infection,donaldson2009mortality,hayward2014comparative}. However, despite its ubiquity, the mechanisms behind bimodality remain elusive. While community network topology has been suggested as an important potential cause \cite{herrera2011multiple,hoen2015epidemic}, the confluence of environmental factors and sporadic interventions make it impossible to isolate the effects of community structure. Additionally, bimodal epidemics have not been observed in previous simulation studies based on real-world mobility networks and disease characteristics [see for example \cite{cauchemez2011role,germann2006mitigation}].

\subsection{summary of methods}
In general terms, our simulation strategy is similar to others in the literature \cite{germann2006mitigation,longini2005containing,halloran2008modeling} and is based primarily on a model which is the subject of our previous publication, calibrated to the Australian demographics and mobility \cite{acemod}. For the present work, we made several minor improvements to the process by which the sample population is generated (these are summarized in the methods section).

We initialize our model by creating stochastically-generated sample populations for each year based on Australian census data from 2006, 2011, and 2016. For each year, we then simulate scenarios in which the pandemic reaches Australia from overseas, continuously seeding the epidemic within 50~km of international airports with probability proportional to the average number of incoming passengers at the corresponding airport. The simulation continues for 180 `days', each with daytime and night-time components during which individuals interact at workplaces and within residential communities, respectively. Here, continuous seeding means we infect people at random in the seeding regions at the beginning of each `day' of the simulation, providing a small but continuous stream of new infections from abroad. 

The disease then spreads in households, neighborhoods, schools, and workplaces based on the disease progression and transmission models validated in our previous work \cite{acemod}. These models are similar to those developed for other countries \cite{chao2010flute,longini2005containing}, with transmission probability dependent on context (location and demographics) and infection status (viral titers and symptom expression). Whenever possible, relative transmission and contact probabilities were derived from field studies \cite{germann2006mitigation,cauchemez2011role}. 

For the results reported here, we set the transmissibility of the disease to achieve a basic reproductive ratio $R_o = 2$ (calibrated for 2006), and held all parameters constant while varying the demographic inputs \cite{diekmann1990definition}. All results were averaged over 5 different sample populations for each year with 30 pandemic instances each, for a total of 150 pandemic instances for each year.  Results for other $R_o$ values, along with standard deviations of incidence, prevalence, and attack rate are shown in Supplemental Figures S1, S2, and S3. [Note: Because of the stochastic nature of our seeding protocol, there is a low probability that the contagion will not spread in the population, giving prevalence values near zero. In our entire set of simulations, this occured a total of three times: once for the 2006 population with $R_o = 2$, once for the 2006 population with $R_o = 1.25$, and once for the 2011 population with $R_o = 1$. For the results presented, these three instances were omitted in the analysis.]

\subsection{results and discussion}

Our results indicate a concerning progression in the population's response to pandemic influenza since 2006. The character of this progression is shown by several metrics in Fig. 1. The incidence (number of newly symptomatic individuals) on each day of our simulated epidemics is shown in Fig. 1(a) and indicates a steadily increasing peak infection rate. The total prevalence (number of symptomatic individuals) in the population as a function of time is shown in Fig. 1(b) and illustrates a corresponding trend in peak prevalence. The accumulated attack rate (total number of people affected) in Fig. 1(c) shows only minor changes in the total proportion of people infected, despite the shifts to larger, earlier epidemic peaks. The progressions of peak day and peak prevalence are given in Fig. 1(d).

\begin{figure}
\centering{
\includegraphics[width = 0.8 \textwidth]{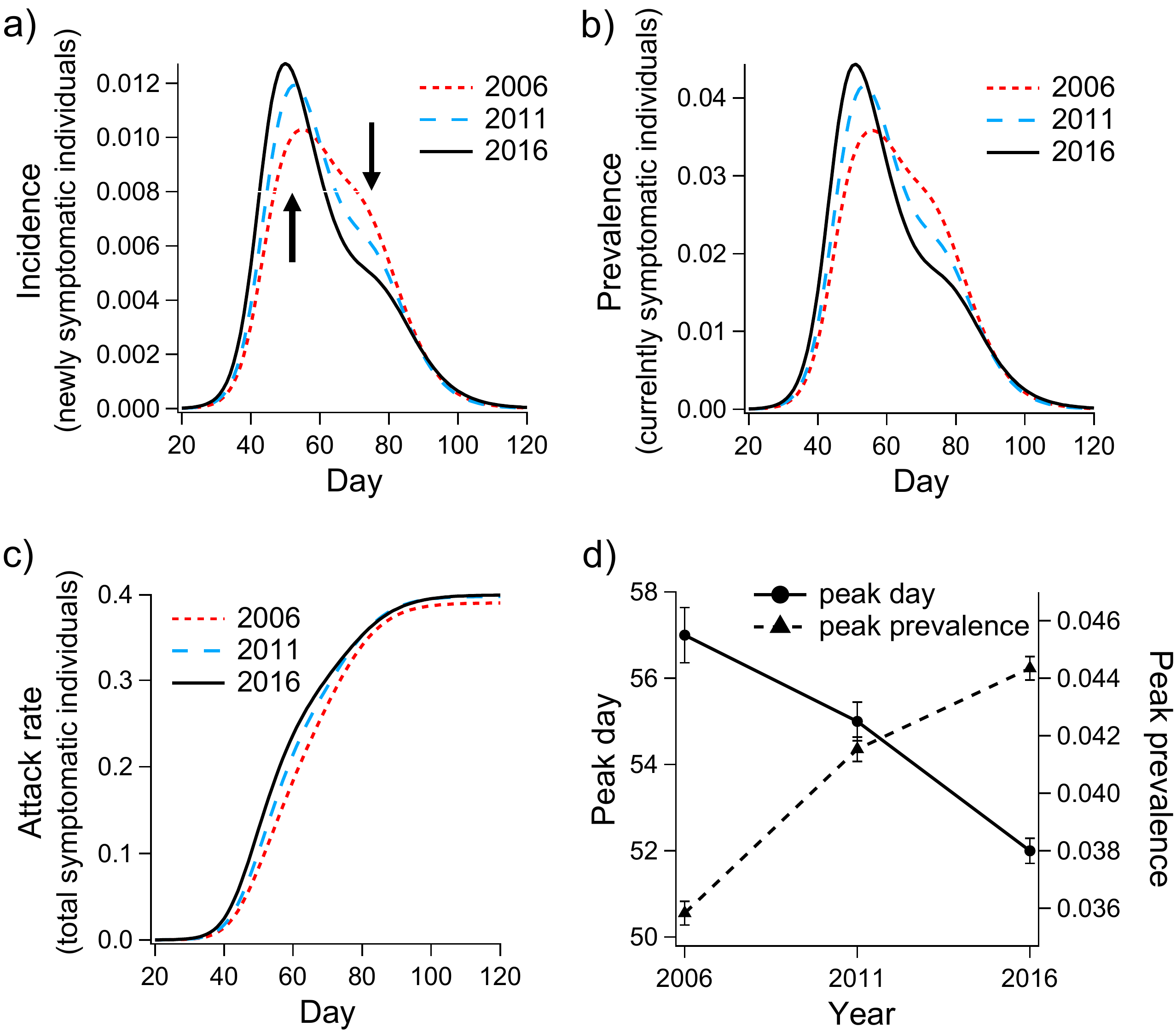} 
\caption{Comparison of 2006, 2011, and 2016 simulation results for influenza incidence (a), prevalence (b), accumulated incidence (attack rate) (c), peak day and prevalence (error bars: $\pm$ S.E.M.) (d). }
\label{result_summary}
}
\end{figure}

At first glance, the stability in overall attack rate [Fig. 1(c)] looks reassuring. This implies that the social network of Australia is not changing in a way that leads to a larger percentage of people becoming exposed to potential infection. A pessimistic but realistic explanation for this is that almost everyone in the simulation is eventually exposed to potential infection in all years. Regardless of the explanation for consistent attack rates, the change in shape of the incidence and prevalence curves leads to some concerning implications. Two of these are illustrated in Fig. 1(d), which traces a correlation between earlier peak days (delay between epidemic onset and peak prevalence) and greater peak prevalence. These results indicate that the virus is spreading more quickly in the population, leading to greater stress on the medical infrastructure and shorter periods of time to respond after the disease is detected. 

For Australia, no empirical estimates exist for the probability of hospitalization given pandemic influenza infection, because the 2009 H1N1 clinical attack rate is unknown. However, based on the documented absolute number of $4,855$ hospitalized cases in 2009, we can estimate the hospitalization probability, given infection, for a reasonable range of attack rate values. Bounding the estimated 2009 attack rate between $10\%$ and $40\%$ translates to hospitalization rates of  $0.22\%$, and $0.056\%$ in the infected population, respectively. Applying these rates to our incidence curves in 2006 and 2016, we estimate the 10-year increase in the number of additional hospitalizations during the peak three weeks of the pandemic to be in the range between 664 and 2,658. Both of these numbers are large enough to contribute significant stress to emergency departments across the country, and would represent a significant increase in hospitalizations due to influenza. Of course, real hospitalization rates depend enormously on the symptom severity associated with the strain, and the levels of pre-existing respiratory health conditions in the population, neither of which are treated explicitly in our model.

%\subsection{bimodality}

In all years, the model output shows strong bimodality in the histogram of local peak-prevalence dates, which corresponds to a transition from an initial wave in urban regions with seeding locations around international airports, and a second wave affecting areas not directly connected to these seed regions (Fig. 2). This bimodality is apparent on the level of Statistical Area level 2 (SA2) regions (see the maps in Fig. 2). In 2006, the initial wave is confined almost exclusively to major cities and the surrounding suburbs. In subsequent years, the trend is an increasing number of rural regions peaking in the first wave. [Note: Statistical Area level 2 is analogous to the Statistical Local Area partitions used in 2006 and earlier years. The hierarchy of partitions used for the 2006, 2011, and 2016 censuses can all be viewed on the ABS website \cite{ABSmaps}.]

\begin{figure}
\centering{
\includegraphics[width=0.7 \textwidth]{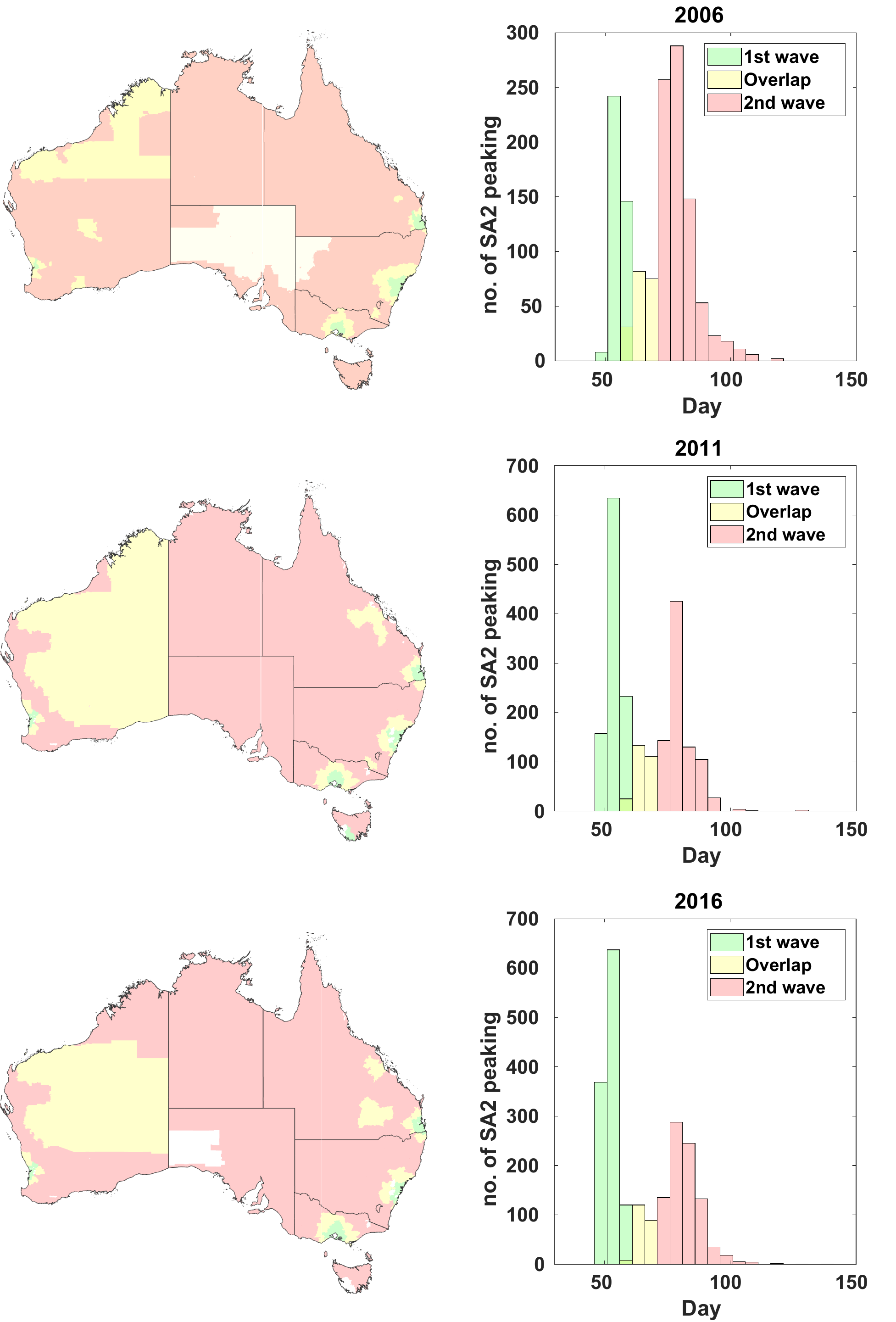}
\caption{Maps and histograms demonstrating geographic and temporal bimodality in epidemic spread. The histograms represent the number of statistical areas (SA2) experiencing peak disease prevalence on a given day. The colors correspond to heuristic classification, green bars indicate the first wave, yellow bars are undetermined, and red bars indicate the second wave. The colors on the map correspond to those in the histogram and demonstrate the geographic distribution of each pandemic wave.}
}
\label{bimodal}
\end{figure}

Bimodality is also detectable on the state level. For example, comparing the 2006 total incidence in New South Wales (which contains the largest international airport, located in Sydney), to that of South Australia (which receives lower levels of international traffic) indicates, unsurprisingly, that states with large international airports tend to contribute to the initial wave, while the states with lower international traffic are affected later and contribute to the second wave [Fig. 3(a)]. The bimodal dynamics are acutely visible in time-series of regional prevalence (please refer to Supplemental Movies 1-3 for dynamic visualizations of disease spread).  

\begin{figure}
\centering{
\includegraphics[width= 0.8\textwidth]{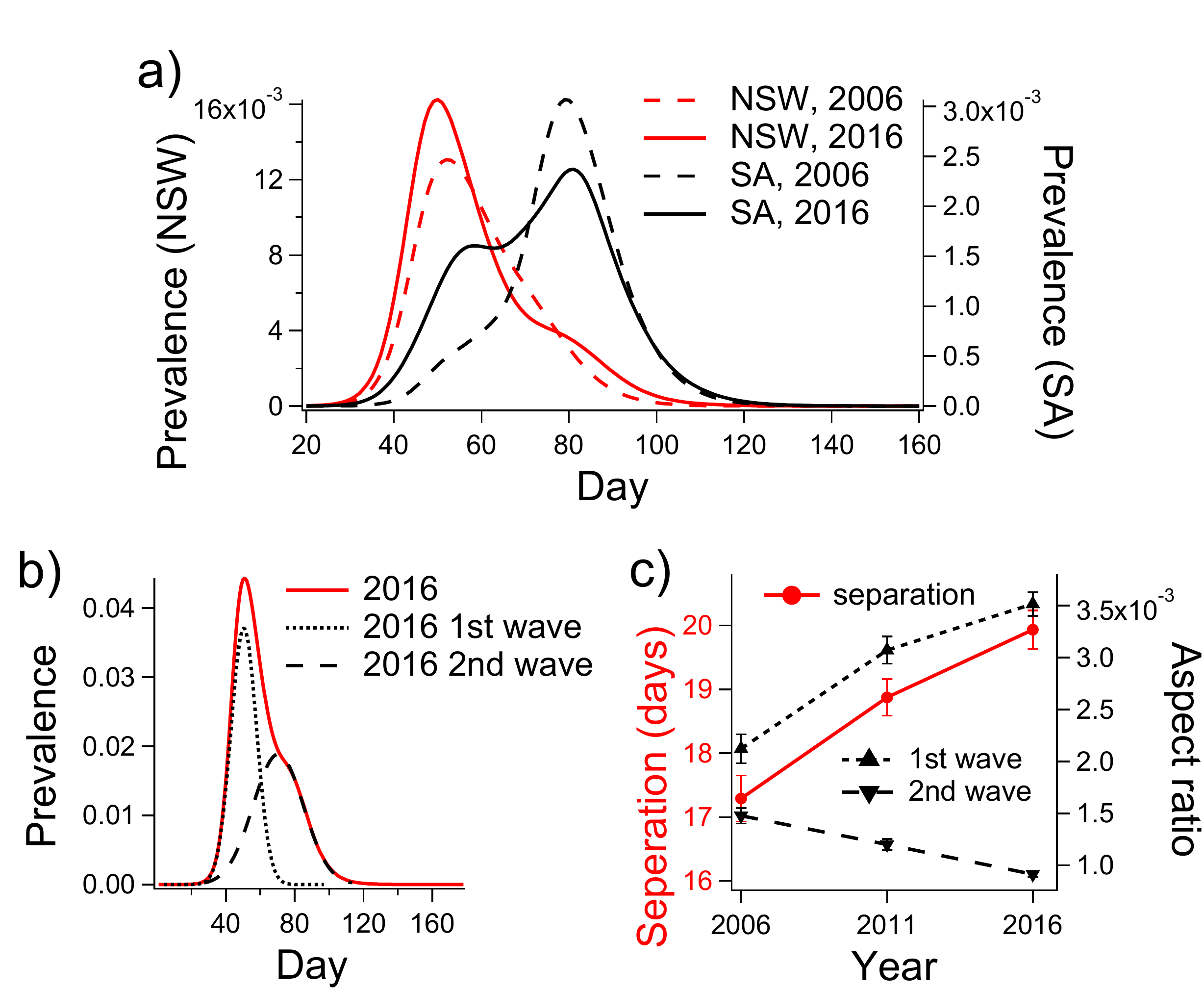}
\caption{Analysis of bimodality in disease prevalence. Plot (a) shows the state-level prevalence for New South Wales (NSW) and South Australia (SA), comparing prevalence curves for 2006 and 2016. Plot (b) shows national prevalence for 2016 and the two Gaussian curves used to fit the data. Plot (c) shows increasing inter-peak separation, increasing aspect ratio of the first mode, and decreasing aspect ratio of the second mode (error bars: $\pm$ S.E.M.). }
\label{peak_analysis}
}
\end{figure}

To quantify the progression of bimodal character which is qualitatively observed in Fig. 2, we fit the prevalence data for each year to pairs of Gaussian curves corresponding to the first and second epidemic waves ($\chi ^2 = [6.63, ~6.62, ~26.7]\times 10^{-6}$ for 2006, 2011, and 2016 respectively) . This fitting procedure is illustrated for 2016 in Fig. 3(b), while the trends in inter-peak separation and peak aspect ratio [the ratio of peak height and standard deviation] are shown in Fig. 3(c). These indicate progressive sharpening of the first wave and broadening of the second wave. This trend corresponds to increased separation of the two peaks, which also manifests as an overall decrease in geographic synchrony: $[0.071, ~0.057, ~0.052 ]~d^{-1}$ for 2006, 2011, and 2016 respectively (quantified here as the reciprocal standard deviations of the histograms of local peak days shown in Fig. 2).

In our assessment, there are two primary mechanisms responsible for these trends. The first is that international air traffic (number of arrivals) increases with time (see Table \ref{arrivals}), which tends to promote the initial wave by introducing a more potent seed infection into the network near urban centers. The second is that the population has become more concentrated in urban regions, close to the international airports where the pandemic is introduced. This trend is clearly observable in plots of population growth comparing urban and rural regions (Fig. 4) \cite{ABSPOP}. The confluence of increased disease influx coupled with a population located closer to the influx point provides a simple but reasonable explanation for the progression in simulated pandemic trends observed across years. 

\begin{figure}
\centering{
\includegraphics[width= 0.9 \textwidth]{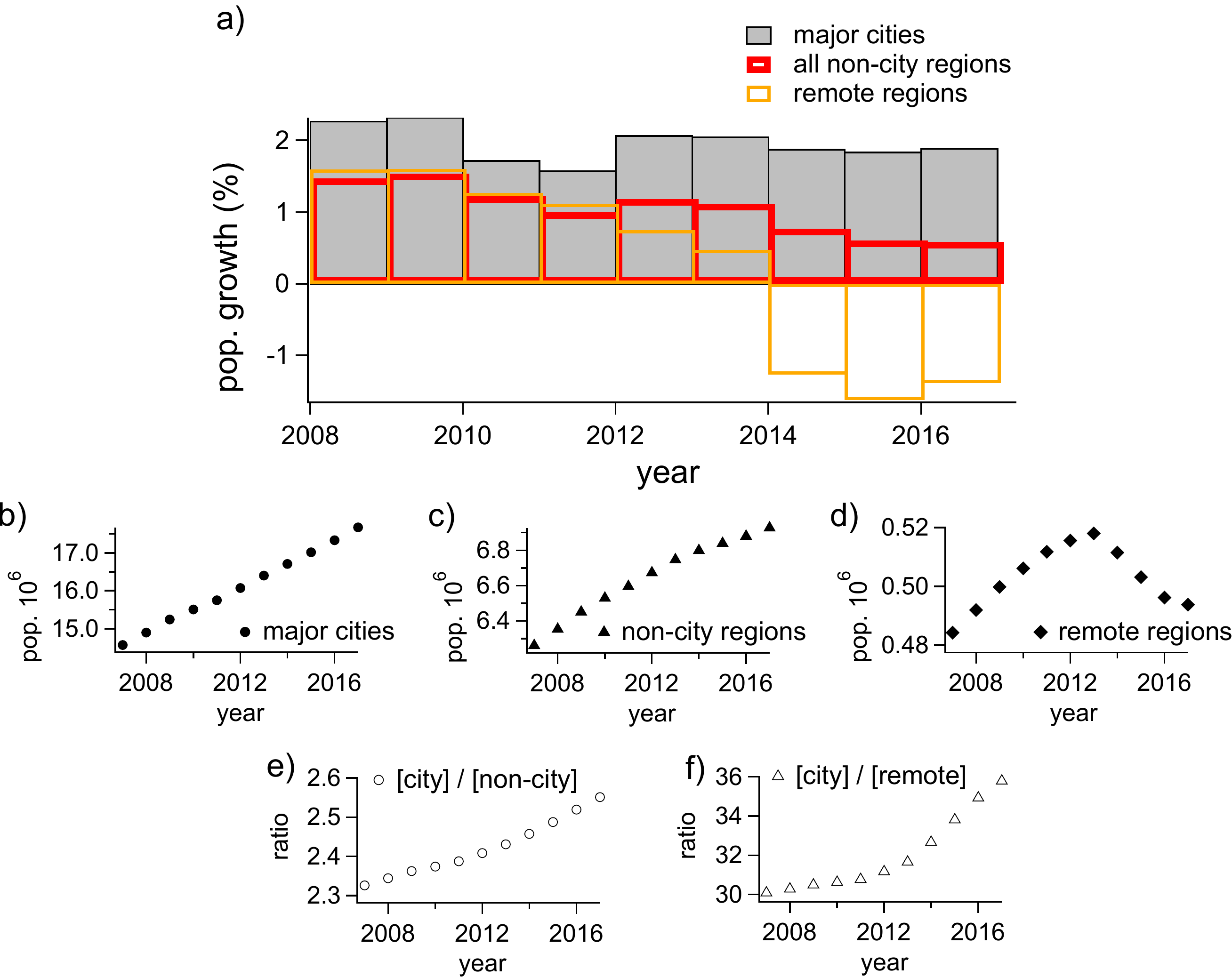}
\caption{(a) Population growth by remoteness. The urban population has been increasing steadily while the rural population declines in both relative (all non-urban areas) and absolute (remote regions) terms. (b - d) Timeseries representations of urban (b), non-urban (c), and remote (d) populations since 2008. (e, f) Timeseries of relative urban population fractions computed against non-urban (e), and remote (f) populations. }
\label{pop_growth}
}
\end{figure}

The role of seeding is demonstrated in Figure 5, which shows results of control studies in which the 2006 and 2011 populations were seeded with 2016 international passenger traffic [Fig. 5 (a, b)]. The residuals between these and the 2016 prevalence levels are shown in Fig. 5(c), and show peaks corresponding approximately to the peaks of the second epidemic waves in both 2006 and 2011 [Fig. 5(c, black and orange dashed lines, respectively)].

\begin{figure}
\centering{
\includegraphics[width=0.8 \textwidth]{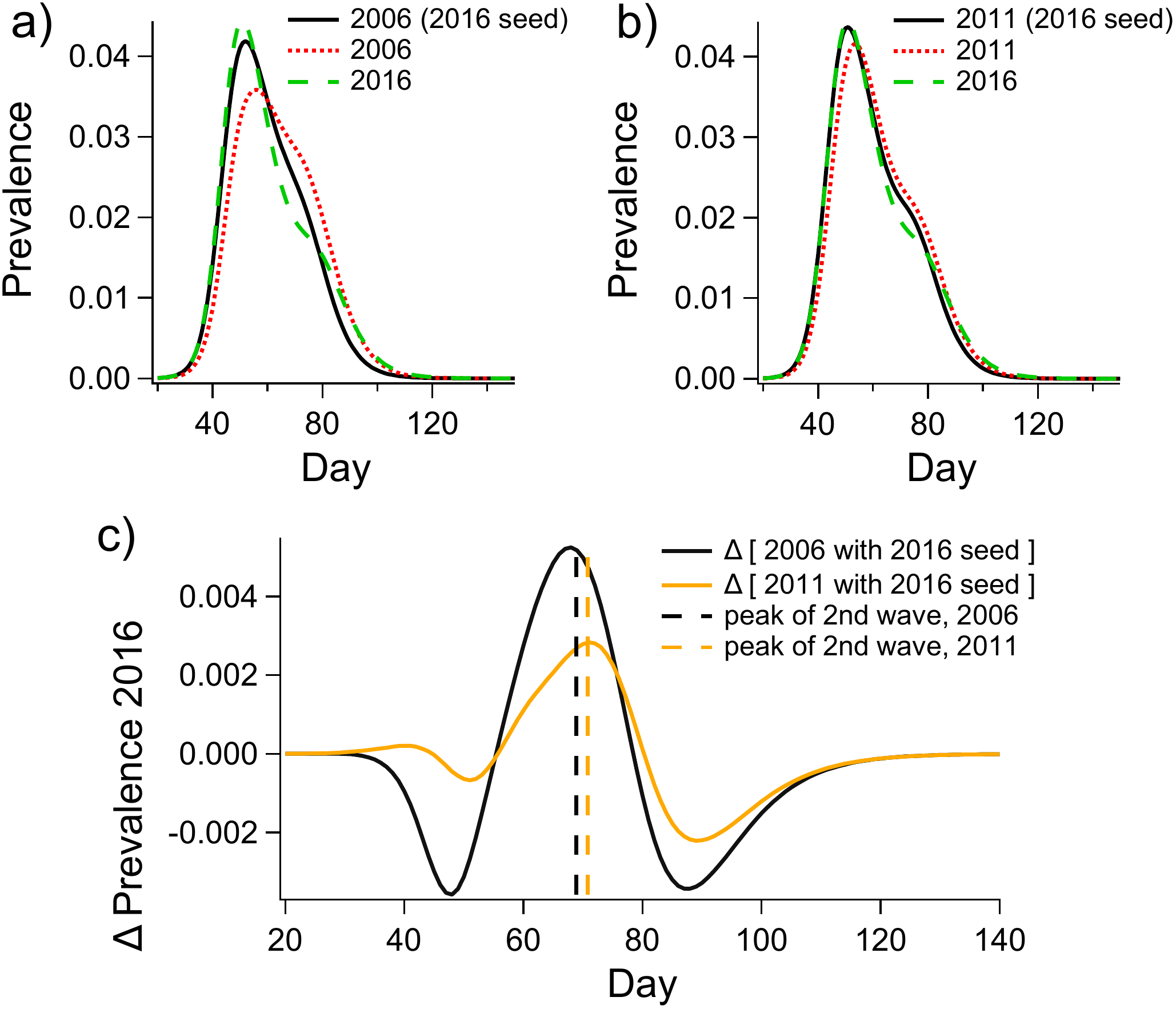} 
\caption{The role of seeding. Plots (a) and (b) show raw prevalence curves demonstrating the effect of applying the seeding conditions of 2016 on the topologies of 2006 (a) and 2011 (b). Plot (c) shows the residual between prevalence in 2016 and prevalence for 2006 (black curve) and 2011 (orange curve) when seeded with 2016 airport traffic. The dashed vertical lines in (c) indicate the maxima of Gaussian fits for the second epidemic wave. }
\label{seed_control}
}
\end{figure}

The coincidence of these residual peaks and the second epidemic waves indicates that the seeding conditions have a larger impact on the first wave than on the second. Indeed, the control for 2011 has an almost identical prevalence profile to that of 2016 during the peak of the epidemic. On the other hand, seeding does not account for the decrease in the intensity of the second pandemic wave from year to year, a trend which we ascribe to increased urbanization. 

In our model, the epidemic can only spread outside of the seeding regions through the work and school commuter networks, which are known to be important to influenza propagation \cite{yang2009transmissibility,Brockmann1337,viboud2006synchrony}. Therefore, the inclusion of rural regions outside of seeding zones in the first wave for 2011 and 2016 pandemics (clearly observable in the maps [Fig. 2]) indicates that the travel-to-work (TTW) network is partially responsible for the increase in the first pandemic wave and decrease of the second. However, the concentration of the urban population within the seeding regions makes this contribution relatively minor because direct seeding and local transmission dominates the dynamics. While it is beyond the scope of the present work, the coupled effect of urbanization trends and mobility network evolution on epidemic spread will be the subject of future investigations. Contact network properties are known to be essential factors in contagion spread \cite{pastor2015epidemic}. It is feasible that contact network evolution could be decoupled from the urbanization process to stall or reverse the observed trends in pandemic dynamics. Through this line of inquiry we hope to design network-based pandemic intervention and urban engineering strategies that will make rapidly urbanizing societies more resilient to pandemics \cite{piraveenan2013percolation}. 

\subsection{conclusion}
By applying a high-fidelity epidemic simulation to of the spread of influenza through the Australian population of 2006, 2011, and 2016 respectively, we have observed an increase in pandemic spreading rate and peak prevalence, combined with decreasing urban-rural bimodality. These trends are associated with increased international air travel and a more urbanized population distribution. The net effect of these two co-acting mechanisms is a shift from a highly bimodal epidemic (both temporally and geographically), to an increasingly mono-modal dynamics where the initial wave dominates. The mobility network extends this phenomenon outside of urban seeding regions. Incidentally, though we observe a net decrease in geographic synchrony over the years investigated here, the observed trends appear to predict a future increase in national epidemic synchrony as the second wave becomes negligible and the initial wave becomes more prominent and peaked. The trend in peak prevalence is particularly worrying as it indicates a non-linear increase in strain on medical infrastructure that current public health policy does not take into account with fixed per capita hospital capacity. Mechanistically, increased air traffic accounts for the trend of earlier peak dates and increased magnitude of the first pandemic wave, while the commuter network and shifting population distribution are responsible for the systematic decrease in the intensity of the second pandemic wave. All three of these causal factors are associated with increased urbanization. To our knowledge, this work provides the first example of a) pandemic simulations across years comparing historical time points, and b) epidemic bimodality mechanistically associated with an empirical interaction network representing a real social system.

\section{detailed methods}\label{meth}

\subsection{data pre-processing}

Census data used to produce the sample populations were pre-processed to remove non-geographical census regions and ensure consistency across different spatial scales.  

Australian census data is publicly available through the ABS' online platform Census TableBuilder. To generate our sample populations we used a subset of demographic data relating to the following characteristics of interest: dwelling location, employment status/work location, household composition, sex, and age. To gather this information, we accessed the following datasets:

\begin{itemize}

\item{SA1 (UR) by AGEP and SEXP}

\item{SA1 (UR) by CDCF and NPRD}

\item{SA1 (UR) by DZN (POW)}

\end{itemize}

[Abbreviations: Statistical Area level 1 (SA1), usual residence (UR), population count by age (AGEP), population count by sex (SEXP), count of dependent children in Family (CDCF), number of persons usually resident in dwelling (NPRD), employment destination zone (DZN), place of work (POW). ]
\\

All exported data is subject to perturbation to preserve the anonymity of individuals as per the Australia Census and Statistics Act 1905. The relative effect of perturbation becomes more prominent with the increased parsing of data. This effect is most severe in the TTW networks, as these break the population into the smallest groups.

The net effect of these perturbations can lead to inconsistencies between data amalgamated for different spatial partitions. For example, accumulating commuter data on the level of (small) SA1 partitions cannot reproduce the ABS-provided statistics over (larger) SA2 partitions, and can lead to the creation of non-existent edges in the SA2 (UR) to SA2 (POW) TTW network. 

To avoid artificial links in the disease transmission network, we removed those SA1 to DZN entries that could not be accounted for on the level of SA2 [i.e., that produced non-existent edges when amalgamated to the scale of SA2 (UR) to SA2 (POW)]. Due to a change in the ABS procedure for introducing perturbations into the 2016 data, some additional processing was required to ensure consistency between 2011 and 2016. This procedure involved the recovery of TTW network edges by sampling over several additional ABS data sets. 

We will release this modified dataset and outline our sampling procedure in detail in our forthcoming publication \cite{ABS_data_paper}.

\subsection{initialization}
We begin by generating sample populations based on census data from the three years investigated. The datasets that inform our sampling procedure describe local-area populations on the order of several hundred people (Statistical Area level 1, and Census District for 2016/2011, and 2006, respectively). These datasets provide population (e.g., age, sex, employment status) and housing (e.g., household size and composition) statistics. These are used as (dependent) probability density functions in the stochastic generation of households and agents respectively. Additional details of the population generation procedure can be found in the supplementary material. We position schools pseudo-deterministically based on their postal code as reported by the Australian Curriculum, Assessment and Reporting Authority (ACARA), a non-census dataset that contains the most complete information available on school enrollment numbers and locations since 2008 (note: we used 2008 school locations and enrollments in place of 2006 data that was not available) \cite{acara}. We then assign students to schools based on the proximity rules described in our previous work \cite{acemod}. 

\subsection{seeding the pandemic} 
In order to realistically model the domestic spread of the disease we assume that the Australian population is exposed to the strain once it is a global pandemic. Following the approach of \cite{germann2006mitigation}, we model this influx of disease by introducing the pandemic to local areas within 50~km of international airports every day. This dynamic seeding procedure infects the population within the seeding zone proportionally to the average daily incoming number of passengers reported by the Bureau of Infrastructure, Transport and Regional Economics (BITRE)\cite{bitre}. Full details of the dynamic seeding procedure can be found in \cite{acemod} Section 3.4. Although the seeding procedure between years is methodologically identical, differences in international airport traffic between years (see table \ref{arrivals}) mean that the dynamic seeding procedures are crucially different.

\begin{center}
\begin{table}[h]
\centering{
\caption{}
\bgroup
\def\arraystretch{1.5}
\begin{tabular}{ |M{2cm}|M{1.25cm}|M{1.5cm}|M{1.5cm}|M{1.5cm}|   }
\hline
\multicolumn{5}{|c|}{Average daily incoming international air traffic }\\
\hline
    Airport & State & \multicolumn{3}{|c|}{ year} \\
\hline
 \multicolumn{2}{|c|}{} & 2006 & 2011 & 2016\\
\hline
Sydney & NSW  & 13214 & 15995 & 19991\\
\hline
Melbourne & VIC  & 5923 & 8557 & 12802\\
\hline
Brisbane & QLD  & 5053 & 5946 & 7299\\
\hline
 Perth & WA  & 2766 & 4512 & 5906\\
\hline
Gold Coast & QLD & 285 & 1044 & 1435\\
\hline
Adelaide & SA & 492 & 766 & 1170\\
\hline
Cairns & QLD & 1186 & 707 & 824\\
\hline
Darwin & NT & 160 & 356 & 355\\
\hline
Townsville & QLD & 0 & 11 & 39\\
\hline
\end{tabular}
\label{arrivals}
\egroup
}
\end{table}
\end{center}

\subsection{disease transmission}

The pandemic can spread within households, neighborhoods, schools, and workplaces according to the transmission probabilities set out in our previous work \cite{acemod}. During each twenty-four hour period, a daytime phase occurs during which infected individuals can spread the infection to others in their working groups comprising approximately 10 others working in the same ``destination zone'' (these groups include classes, grades, and schools for teachers and students, which have more than 10 people, see \cite{acemod}). This is followed by a night-time phase during which the infection can spread within households, household clusters, and neighborhoods. Importantly, the pandemic can only spread between statistical areas during the daytime phase as this is the largest geographical area of interaction in the night-time phase. Further details of the transmission model are given in the supplemental material.

\section*{Acknowledgments}
The Authors are grateful to Tim Germann, Joseph Lizier, Philippa Pattison, E. Yagmur Erten, Manoj Gambhir, and Stephen Leeder for helpful discussions on agent-based simulation of pandemic influenza. \textbf{Author Contributions:} M. Prokopenko and M. Piraveenan conceived of the project; C. Zachreson, K. Fair, N. Harding, and M. Prokopenko composed the manuscript; C. Zachreson and K. Fair performed the simulations and analyzed the data; N. Harding and O. Cliff assisted with modelling and data pre-processing. All authors contributed to discussions regarding interpretation of results. \textbf{Data availability:} All data needed to evaluate the conclusions in the paper are present in the paper and the Supplementary Materials. Additional data available from authors upon request. \textbf{Competing interests:} The authors declare no competing interests. \textbf{Funding:} The Authors were supported through the Australian Research Council Discovery Project DP160102742.

\bibliography{Year_Comparison_refs}

\bibliographystyle{Science}

\section*{Supplemental material}

%Here you should list the contents of your Supplementary Materials -- below is an example. 
%You should include a list of Supplementary figures, Tables, and any references that appear only in the SM. 
%Note that the reference numbering continues from the main text to the SM.
% In the example below, Refs. 4-10 were cited only in the SM.     

\newcommand{\beginsupplement}{%
 \setcounter{table}{0}
   \renewcommand{\thetable}{S\arabic{table}}%
     \setcounter{figure}{0}
      \renewcommand{\thefigure}{S\arabic{figure}}%
      \setcounter{page}{1}
      \renewcommand{\thepage}{S\arabic{page}} 
     }

%
%\section*{Supporting Information for\\ `Vulnerability to pandemics in a rapidly urbanizing society'}
%\centerline {by Zachreson \it et al.}
%\hspace{1cm}

\beginsupplement
\baselineskip24pt

\subsection*{Supplemental Movies}\label{supmov}

Supplemental movies 1 - 3 show the dynamic progression of influenza in Australia as simulated by our model in 2006, 2011, and 2016, respectively. In addition to the national dynamics, local dynamics for three major cities are shown as well (Sydney NSW, Melbourne VIC, and Adelaide SA). Red circles are centered on the international airport of each city and represent the 50~km radius used for seeding.  

\

\subsection*{Supplemental Figures: Incidence, Prevalence, and Attack rate for different values of $R_o$}\label{supfigs}

Supplemental Figures S1 - S3 show Incidence, Prevalence, and Attack rate (respectively) for values of $R_o$ ranging from 1 to 2. Colored bands represent $\pm$ standard deviation over the 150 simulation instances used for each curve. 

\begin{figure*}
\centering
\includegraphics[width= 0.6\textwidth]{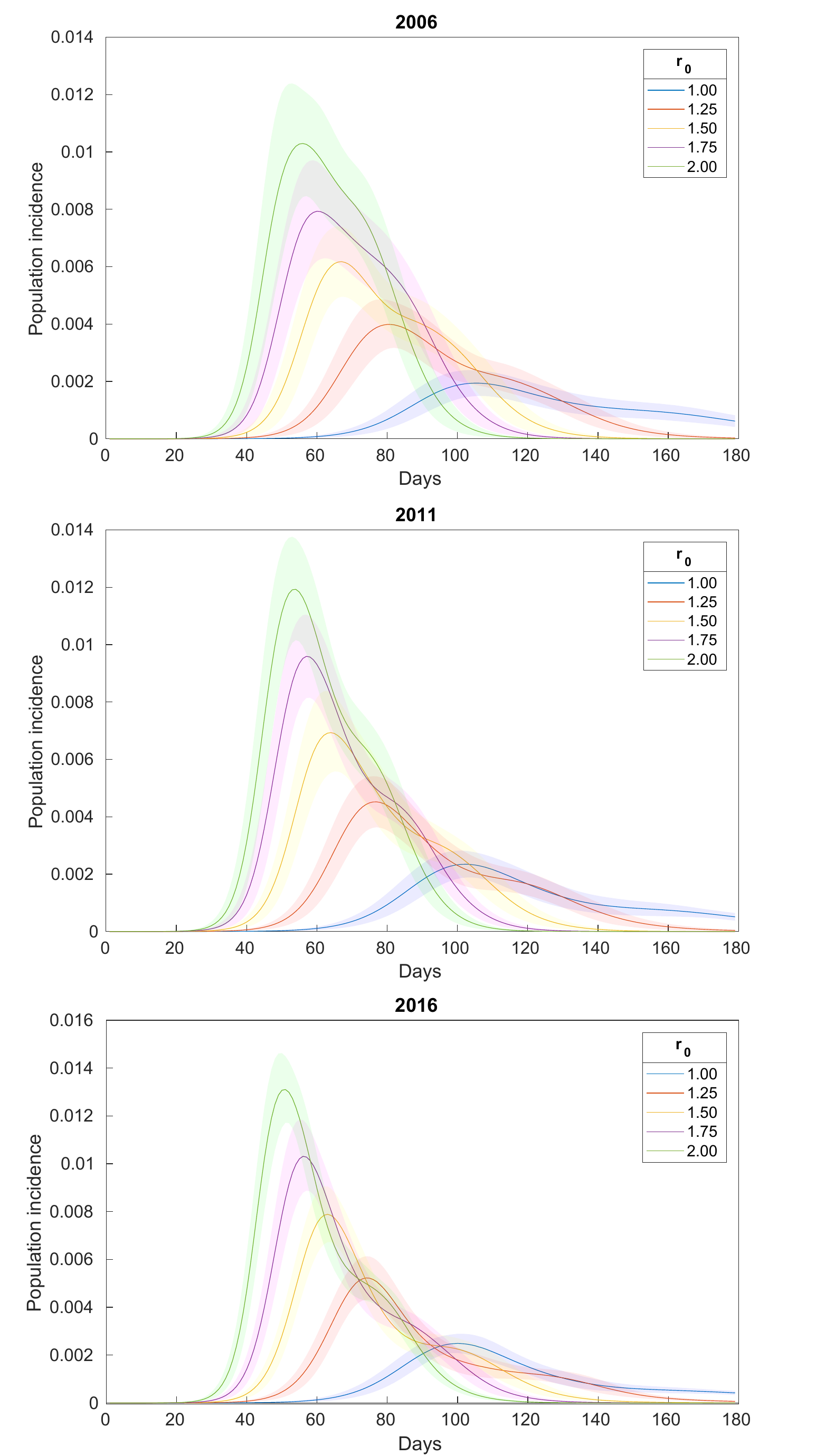}
\caption{Incidence proportion for various $R_o$ values. The curves represent the average simulated pandemic incidence (proportion of newly symptomatic individuals) as a function of time for 2006, 2011, and 2016 as computed $R_o = [1.00, 1.25, 1.50, 1.75, 2.00]$ (green, purple, yellow, red, and blue curves, respectively). The shaded regions around each curve represent $\pm$ standard deviation at each timepoint over 150 instances.}
\label{sup_incidence}
\end{figure*}

\begin{figure*}
\centering
\includegraphics[width=0.6 \textwidth]{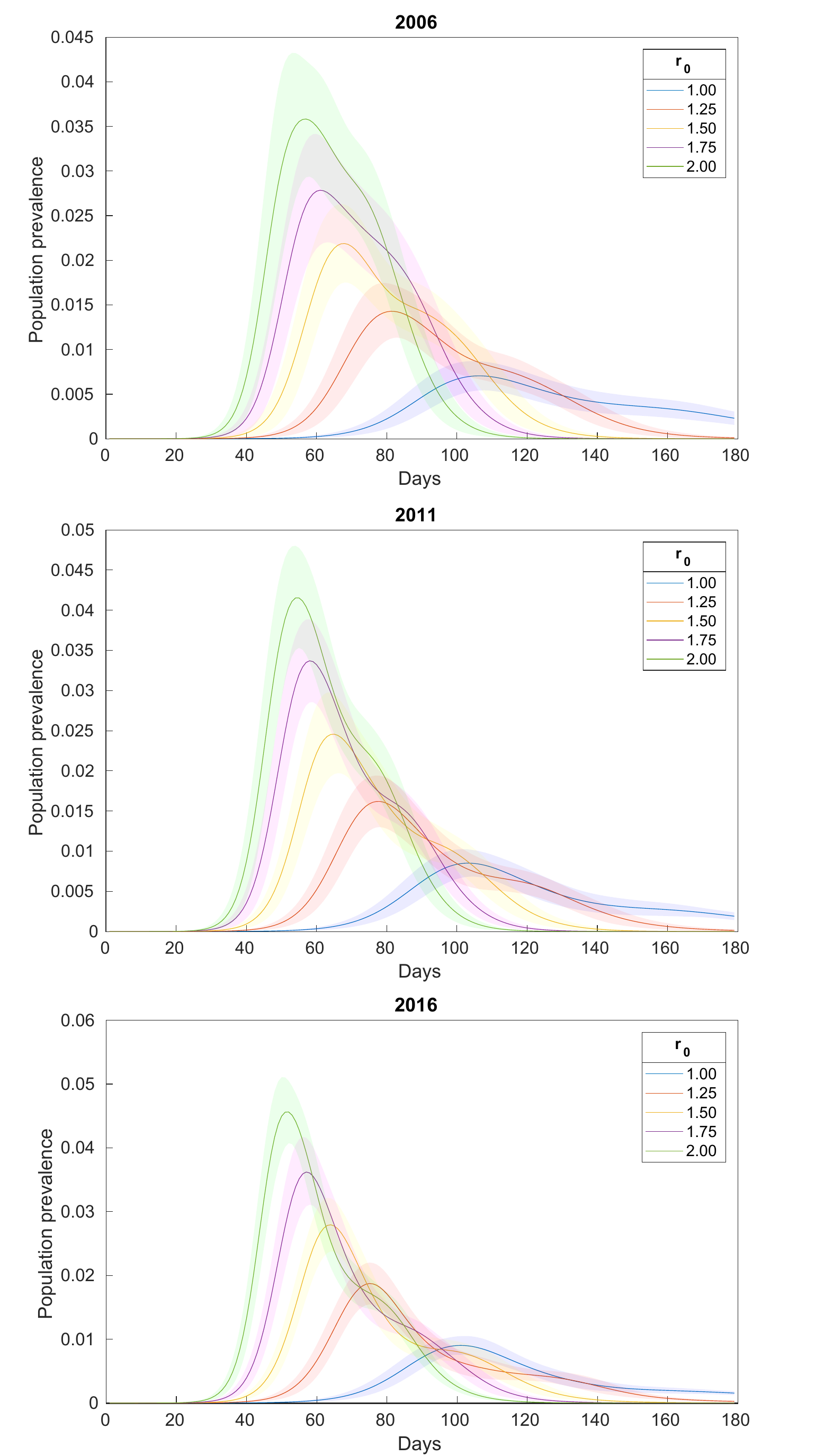}
\caption{Prevalence proportion for various $R_o$ values. The curves represent the average simulated pandemic prevalence (proportion of currently symptomatic individuals) as a function of time for 2006, 2011, and 2016 as computed $R_o = [1.00, 1.25, 1.50, 1.75, 2.00]$ (green, purple, yellow, red, and blue curves, respectively). The shaded regions around each curve represent $\pm$ standard deviation at each timepoint over 150 instances.}
\label{sup_prevalence}
\end{figure*}

\begin{figure*}
\centering
\includegraphics[width=0.6 \textwidth]{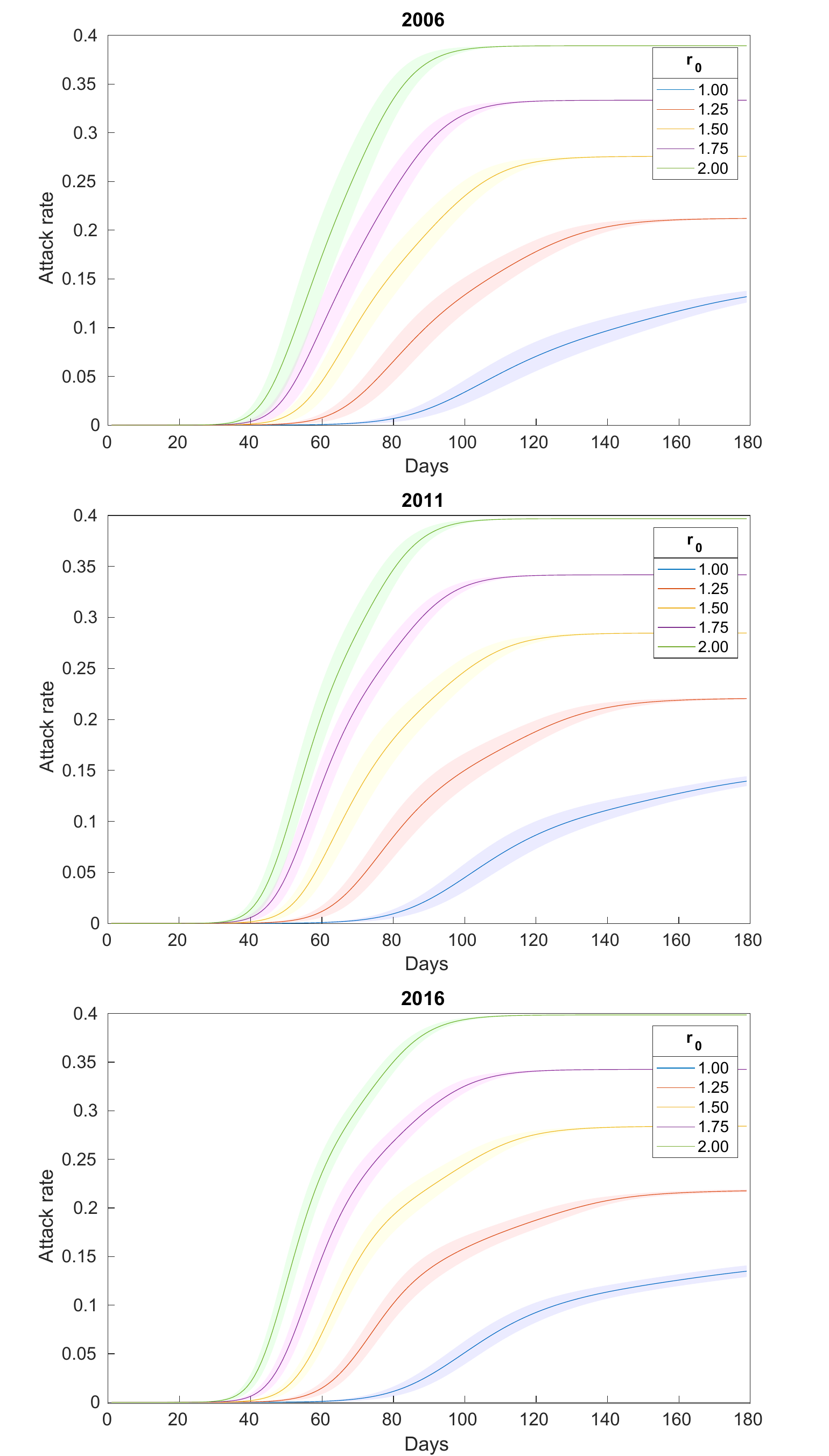}
\caption{Attack rate for various $R_o$ values. The curves represent the average simulated pandemic attack rate [total proportion of symptomatic individuals (accumulated incidence) for the entire pandemic up to time $t$] for 2006, 2011, and 2016 as computed $R_o = [1.00, 1.25, 1.50, 1.75, 2.00]$ (green, purple, yellow, red, and blue curves, respectively). The shaded regions around each curve represent $\pm$ standard deviation at each timepoint over 150 instances.}
\label{sup_attack_rate}
\end{figure*}

\pagebreak

\subsection*{Model Description}

\subsubsection*{Demographics and mobility}
We utilize the ABS census data from 2006, 2011, and 2016 to distribute the sample Australian population into geographical areas on the scale of Statistical Area Level 1 (SA1), for 2011 and 2016 censuses, and the equivalent scale of Census District (CD) for the 2006 census.  These areas correspond to geographical regions containing 200-800 residents. This is the highest resolution at which the ABS provides population~(e.g., age, sex, employment status) and housing~(e.g., household size and composition) statistics. We use these statistics as (dependent) probability distributions in the generation of sample populations.

The stochastic generation of a sample population starts by iterating through SA1s. For a given SA1, the agent generation process begins by generating a household type. The household types are lone, group, or family types. These are then further subdivided into the following categories: single, couples with children, couples without children, or single parent families. This household is then assigned a size, conditional on its type, and agents are created corresponding to the household type with attributes such as age and sex. This process continues until the chosen SA1 has as many agents as its known population after which new SA1s are chosen until all are fully populated.

After the agent population has been generated for all SA1s, workplaces and schools are assigned to adults and children respectively. Workplaces are assigned based on the ``Travel to work" (TTW) matrix which gives a home SA1, a workplace DZN and a number of individuals who live and work in these locations. As the home regions for all agents have already been defined, agents are assigned to workplaces to satisfy the TTW data exactly. 

Unfortunately the ABS census does not contain data about the location or attendance of schools. As such, the schools were placed based on data from ACARA. The placement procedure requires a reasonable correspondence between postcode and Statistical Area Level 2 (SA2), which have different partition schemes, but exist on approximately the same geographical scale. We established correspondence between postcodes and SA2s based on the following rules: correspondence between a postcode-SA2 pair is identified if at least one of the following conditions is met:

\begin{itemize}
\item{at least $10\%$ of the postcode area is encapsulated by the SA2}
\item{at least $10\%$ of the SA2 area is encapsulated by the postcode}
\end{itemize}

In cases where multiple SA2s correspond to the same postcode, we assign schools in that postcode to SA2s based on the population distribution of student-aged individuals in the candidate SA2s. This is done iteratively, that is, after a school is placed, the local student populations (with respect to further school placements) are reduced by the size of the school assigned. Using this method, schools end up distributed based on both their documented postcode, and also the local distribution of student-aged individuals. 

In the absence of direct data regarding school attendance, we stochastically assign students to schools based primarily on proximity, assuming that students would go to a school if they live within the approximate catchment zone of the school. If a student is within the catchment zone of multiple schools, the school is chosen randomly from these based on the number of available places. Full details of this assignment can be found in our previous work ({\it31}).

\subsubsection*{Natural History of Disease}
The natural history of a disease describes the course of a disease from onset to resolution. This description includes a number of states which describe the properties~(e.g., infectious or non-infectious) depending on the disease being modeled. The model of influenza used in this paper has 5 distinct states: \textsc{susceptible}, \textsc{latent}, infectious \textsc{symptomatic}, infectious \textsc{asymptomatic} and \textsc{recovered}. The \textsc{latent} period captures the time from initial exposure to the pathogen to the onset of infectiousness. From the \textsc{latent} state, individuals progress to either \textsc{symptomatic} or \textsc{asymptomatic} in which case they are half as infectious as \textsc{symptomatic} individuals. An individual that will eventually show symptoms of influenza will typically do so a number of days after contracting the disease. Therefore, before expression of symptoms, these agents are considered to be in the infectious \textsc{asymptomatic} state. At the conclusion of their infectious period~(\textsc{symptomatic} or \textsc{asymptomatic}), individuals progress to the \textsc{recovered} state in which they are not infectious and are immune to infection. The parameters for the natural history of disease used in this model results in generation times of between 3.35 and 3.39 days dependent on the value of $R_0$. Full details of the natural history of disease can be found in our previous work ({\it31}).

\subsubsection*{Transmission model}
In this model, infected individuals spread disease to others via a number of mixing groups. During the daytime phase individuals can spread the
infection to others in their working groups comprised of approximately
10 others working in the same “destination
zone” (DZN). For teachers and students, these working groups include classes, grades, and schools. For each individual $i$ there is set  of mixing groups $\mathcal{G}_i(n)$ which that individual interacts with. Each mixing group $g \in \mathcal{G}_i(n)$ is associated with a static set of agents $\mathpzc{A}_{g}$.

At an individual level we are interested in the probability of a susceptible agent i becoming infected during a given
time period (step) $n$. Let $X_i(n)$ denote the state of individual $i$ at time $n$. The probability of a susceptible agent $i$ becoming infected during a given time period is given by 
\begin{equation}
p_i(n) = Pr\{X_i(n) = \textsc{latent} | X_i(n-1) = \textsc{susceptible}\}
\end{equation}
where \textsc{latent} denotes the first stage of infection.
The infection probability of a susceptible individual $i$ is computed as 
\begin{equation}
p_i(n) = 1-\prod_{g \in \mathcal{G}_i(n)} \left[ \prod_{j \in \mathpzc{A}_{g} \setminus i}\left( 1-p^g_{j \rightarrow i}(n) \right)  \right]
\end{equation}

where $\mathcal{G}_i(n)$ denotes the mixing groups which individual $i$ interacts at time step $n$,  $p^g_{j \rightarrow i}$ denotes the instantaneous probability of transmission from agent $j$ to agent $i$ in contact group $g$. The context-based probabilities of transmission are taken directly from studies where possible. Where context based transmission probabilities were not available, additional steps were taken to calculate transmission probabilities from known contact rates. Full details can be found in our previous work ({\it31}).

\end{document}